\documentclass[amsmath,amssymb,preprint]{revtex4-1}
\usepackage{dcolumn}
\usepackage{bm}

\usepackage{graphicx}
\usepackage{epstopdf}

\usepackage{textcomp} 
\newcommand{\Si}{Si$(111)-(7\times7)$ }
\newcommand{\BiTe}{BiSbTe$_3$ }
\usepackage{enumitem}
\begin{document}

\title{Four-point probe measurements using current probes with voltage feedback to measure electric potentials}
\author{Felix L\"upke}

\author{David Cuma}

\author{Stefan Korte}

\author{Vasily Cherepanov}

\author{Bert Voigtl\"ander}
\altaffiliation[Electronic mail: ]{\urlstyle{same}\url{b.voigtlaender@fz-juelich.de}}
\affiliation{Peter Gr\"unberg Institut (PGI-3), Forschungszentrum J\"ulich, 52425 J\"ulich, Germany}
\affiliation{JARA--FIT, 52425 J\"ulich, Germany}

\date{\today}

\begin{abstract}
We present a four-point probe resistance measurement technique which uses four equivalent current measuring units, resulting in minimal hardware requirements and corresponding sources of noise. Local sample potentials are measured by a software feedback loop which adjusts the corresponding tip voltage such that no current flows to the sample. The resulting tip voltage is then equivalent to the sample potential at the tip position. We implement this measurement method into a multi-tip scanning tunneling microscope setup such that potentials can also be measured in tunneling contact, allowing in principle truly non-invasive four-probe measurements. The resulting measurement capabilities are demonstrated for \BiTe and \Si samples.
\end{abstract}
\maketitle

\section{Introduction}
Four-terminal sensing and four-point probe methods are widely used measurement techniques which allow to determine the electrical resistance of a sample \cite{Schroder2006}. In contrast to classical two-terminal measurements, this technique eliminates lead and contact resistances from measurements and thereby allows the precise measurement of resistances with high accuracy. Originally proposed by F. Wenner for application in geophysics \cite{Wenner1915}, today four-probe measurements can be found in a wide range of applications including semiconductor characterization \cite{Schroder2006}. Due to their high accuracy, four-probe measurements are often used in fundamental research such as (low temperature) transport measurements where the sample under investigation is typically contacted by lithographic contacts \cite{Jaschinsky2008}, monolithic four-point probes \cite{Hofmann2009,Dangelo2009,Thorsteinsson2009,Petersen2010} or a multi-tip scanning tunneling microscope (STM) \cite{Hobara2007,Martins2014,Just2015,Luepke2015,Luepke2017,Luepke2017a}. In general, the four-probe measurement setup consists of two biased contacts which inject a current into the sample and two contacts which measure the resulting voltage drop across the sample, e.g. by high ohmic volt meters \cite{Schroder2006,Hobara2007,Just2015}. The four-probe resistance of the sample is then determined from the slope of the resulting four-probe $I/V$ curve \cite{Hobara2007}.
\begin{figure}[!b]
\includegraphics{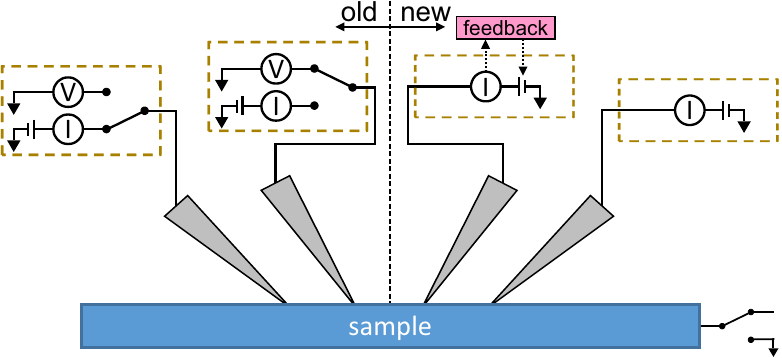}
\caption{Schematic of a four-point measurement with tips of a multi-tip STM forming the contacts on the sample surface. (left) A measurement setup, where each tip is connected to identical electronics which can be switched between voltage and current measurement mode by a relay. (right) A setup, where each tip is connected to identical biased current measurement electronics and voltage measurements are performed via feedback loops. The complete four-probe measurement setup results by mirroring at the central vertical dashed line, respectively. In both realizations, the sample ground contact is connected to approach the tips to the sample and is disconnected during the four-probe measurements.}
\label{0}
\end{figure}

Previously reported implementations of four-probe measurements into a multi-tip STM require two different kinds of electronics for current and voltage sensing, respectively \cite{Hobara2007}. As a result, in order to approach the tips to the sample surface and to exchange contacts during four-probe measurements, the tips have to be rewired to the different electronics in between measurements. One approach to simplify this process is to use four instances of identical electronics which can perform both, voltage and current measurements \cite{Hobara2007,Luepke2015} as shown in Fig. \ref{0} (left). The switching between the two measurement modes is then performed e.g. by a relay \cite{Hobara2007,Luepke2015}. A drawback of this approach is, that it leads to bulky, complex electronics and introduces additional noise to the measurements due to the additional circuitry.

An alternative way to measure local sample potentials in a STM is the scanning tunneling potentiometry method \cite{Muralt1986,Druga2010,Luepke2015}. In this method, the voltage applied to the tip is controlled by a feedback loop, such that the current to the sample vanishes, i.e. the electrical potential of the tip and the sample at the position of the tip are identical. The voltage resulting in a vanishing current is then recorded as the local electric potential of the sample.
\begin{figure*}
\includegraphics{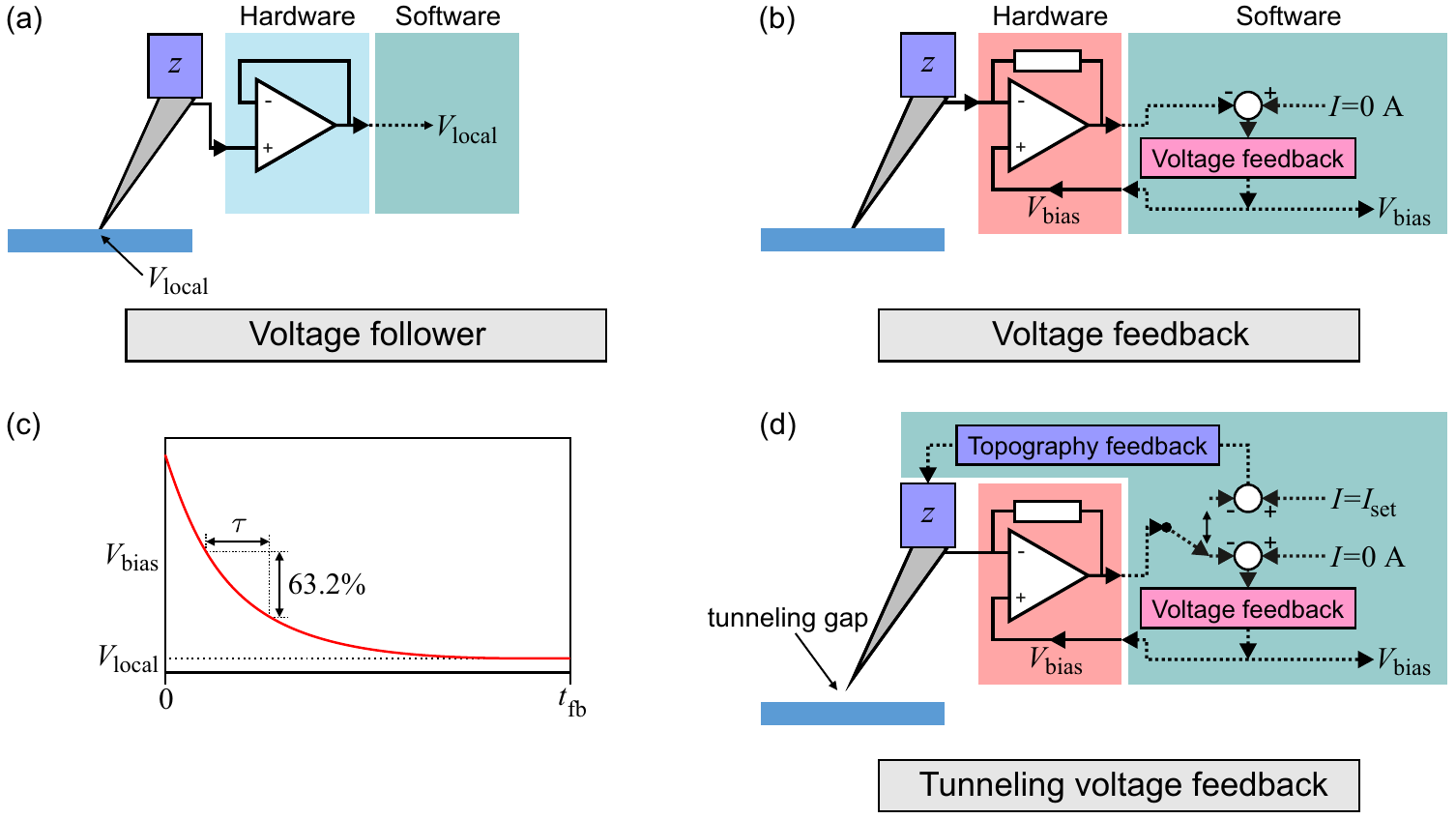}
\caption{Schematic of the three voltage measurement techniques. Solid lines correspond to hardware connections and dashed lines to software implementations. At the interfaces of hardware and software realizations, analog-to-digital and digital-to-analog converters transform the signals accordingly (not shown). (a) Schematic of the voltage follower implementation with the tip in direct contact with the sample surface. The local sample potential $V_{\rm local}$ is directly passed to the output of the voltage follower circuit and into the software. (b) Schematic of the software voltage feedback implementation which continuously adjusts the current through the tip to $0\rm\,A$ (such that $V_{\rm bias}=V_{\rm local}$) in direct contact to the sample surface. (c) Schematic of the working principle of the voltage feedback. The tip voltage $V_{\rm bias}$ is adjusted to the local sample potential $V_{\rm local}$ for a feedback duration $t_{\rm fb}$. The bandwidth of the feedback loop is $\Delta f$, which we determine as the decrease of $V_{\rm bias}-V_{\rm local}$ by $1-e^{-1}\approx63.2\%$ in the time interval $\tau=1/(2\pi\Delta f)$. (d) Schematic of the tunneling voltage feedback implementation in which the software alternatingly performs topography feedback (setpoint $I=I_{\rm set}$) and voltage feedback (setpoint $I=0\rm\,A$) on the timescale of $\rm ms$.}
\label{2}
\end{figure*}

Here, we present a four-probe measurement implementation where voltage measurements are performed by means of current measurement in combination with a software voltage feedback loop. This approach requires only four identical, minimalistic, biased current measurement units, as hardware voltage measurement circuitry and relays are redundant. In the resulting setup, shown in Fig. \ref{0} (right), current and voltage probes can be flexibly exchanged without any rewiring. Furthermore, the voltage sensing can be performed in tunneling contact such that truly non-invasive four-probe measurements become possible which is advantageous e.g. for the characterization of fragile samples.

This paper organizes as follows. In Sec. \ref{methods}, we focus on one of the voltage probes where we discuss different kinds of realizations of the voltage measurement. In Sec. \ref{noise}, we determine the theoretical noise levels of the different voltage measurement realizations and subsequently the entire four-probe measurement setup. In Sec. \ref{results}, we demonstrate four-probe measurement capabilities on \BiTe and \Si samples under ultra-high vacuum conditions. In Sec. \ref{appl}, further applications and improvements of the present setup are discussed.

\section{Methods} \label{methods}
The experimental setup used here is a home-built room temperature four-tip STM with electrochemically etched tungsten tips. Further details on the experimental setup can be found in Refs. \onlinecite{Cherepanov2012} and \onlinecite{mProbes2015}. In this setup, we compare different four-probe measurement methods in which the voltage measurement is realized in three different ways (A) to (C). 

\subsection{Voltage follower}
In this voltage measurement method, the voltage sensing tip is connected to a typical voltage follower circuit, as used in previous setups \cite{Hobara2007} and as shown schematically in Fig. \ref{2} (a). The tip is brought into direct contact to the sample by approaching it towards the sample surface by a few $\rm nm$ out of the tunneling contact as described elsewhere \cite{Polley2012,Hobara2007}. The resulting junction resistance between the tip and the sample typically is in the range of $\rm k\Omega$ to $\rm M\Omega$. The local sample voltage at the position of the tip $V_{\rm local}$ is measured by a high input-impedance buffer circuit before the signal is read into the measurement software via an analog-to-digital converter (ADC). Between the voltage follower circuit and the ADC a low-pass filter can be used to adjust the measurement bandwidth $\Delta f$ in order to reduce the measurement noise.

\subsection{Voltage feedback}
This approach is based on the measurement of the voltage by nulling the current flowing between the tip and sample, similar to standard methods for measuring voltages with high precision \cite{Abramowitz1967}. As in the voltage follower application, the voltage sensing tip is in direct contact to the sample surface but is only connected to a biased transimpedance (current) amplifier. A schematic of the setup is shown in Fig. \ref{2} (b). The voltage sensing is performed by a proportional-integral software feedback loop, similar to the topography feedback loop typically used in STM. For this purpose, the amount of current through the tip at a certain tip bias voltage $V_{\rm bias}$ is read into the measurement software via an ADC converter and the feedback loop nulls the tip-sample current by continuously adjusting $V_{\rm bias}$ via a digital-to-analog converter (DAC) connected to the $V_{\rm bias}$ input of that tip, until $V_{\rm bias}=V_{\rm local}$ \cite{Druga2010,Luepke2015}. The measurement noise bandwidth in this implementation is given by the bandwidth of the feedback loop, which we determine as the time interval $\tau=1/(2\pi\Delta f)$ in which the difference between currently applied tip voltage and target local sample voltage $V_{\rm bias}-V_{\rm local}$ decreases by $1-e^{-1}\approx63.2\%$, similar to an RC circuit (Fig. \ref{2} (c)). $\Delta f$ can be determined by applying a sine signal to the input of the closed feedback loop and tuning its frequency until a damping of $-3\,\rm dB$ is observed. At this point the sine frequency corresponds to $\Delta f$. Note that the resulting feedback duration $t_{\rm fb}$ which is required for the feedback loop to converge below the noise threshold is typically larger than $\tau$ and depends on the initial difference of the two signals $V_{\rm bias}-V_{\rm local}$.

\subsection{Tunneling voltage feedback}
This method is a special case of (B), where the voltage sensing tip is in tunneling contact rather than direct contact. A challenge in the measurement of the local sample voltage in tunneling contact is that, both the topography feedback and voltage measurement are entangled by their concurrent use of the tunneling current as the process variable \cite{Luepke2015}. One solution for this problem is the temporary stopping the topography feedback during the measurement of the local sample potential. However, a deactivation of the topography feedback results in the tip drifting towards or away from the sample on the time scale of less than a second (at room temperature), resulting in a significant change of the junction resistance, as the tunneling current depends exponentially on the tip-sample distance. Therefore, for typical four-probe measurement durations of up to several minutes it is difficult to hold the tip in tunneling contact throughout the measurement. For this reason, also the measurement of the local sample potential in tunneling contact with a voltage follower is impractical.

The alternating feedback technique \cite{Druga2010,Luepke2015} overcomes this problem. In this method, the STM control software performs alternatingly topography and voltage feedback on the timescale of $\rm ms$. The corresponding schematic setup is shown in Fig. \ref{2} (d). In detail, the tip height in tunneling contact is controlled such that a current set point $I_{\rm set}$ is maintained at a fixed tunneling voltage $V_{\rm bias}=V_{\rm t}$. Then, for a time $t_{\rm fb}$, the topography feedback is stopped (the tip is held at fixed height above the sample), while the tip voltage $V_{\rm bias}$ is adjusted such that the tunneling current vanishes as described in the voltage feedback method and shown in Fig. \ref{2} (c). Afterwards the tunneling voltage $V_{\rm bias}=V_{\rm t}$ is restored, the topography feedback is re-enabled and the procedure is repeated. A drawback of this method is that the switching between the two feedback loops takes up additional time in comparison to methods (A) and (B). As a result, the fundamental noise level for the same measurement duration increases in comparison to a measurement without alternating feedback. However, the striking advantage of this method over previous methods is that the local sample potentials can be measured non-invasively and with the same minimalistic electronics as in the voltage feedback implementation (B).

An alternative way to disentangle the topography and potential feedback in tunneling contact include the use of a DC current component and an AC current component and use them individually for the voltage and topography feedback, respectively \cite{Muralt1986,Pelz1989,Bannani2008}. However, this method requires additional circuitry and is reported to be prone to cross-talk between the AC and DC signal \cite{Nakamura2016}. For this reason, we focus here only on the performance of the voltage follower and alternating feedback techniques.\\

For comparison of the different measurement techniques (A) through (C) we connect each of the four tips to one of four instances of equivalent electronics including a relay which can switch between a voltage follower circuitry (based on OPA111) and a biased commercial FEMTO DLPCA-200 variable gain transimpedance amplifier \cite{Femto2015} to measure the currents with adjustable measurement range. Typically, the gain of the current injecting tips is $10^6\rm\,V/A$, while that of the voltage sensing tips in the voltage feedback techniques is $10^9\rm\,V/A$. For the voltage feedback techniques, we further use voltage dividers of up to $1/1000$ in front of the bias voltage connector of the voltage sensing tip's electronics, in order to increase the voltage feedback accuracy which can otherwise be limited by the minimum step size of the DAC and noise in the $V_{\rm bias}$ signal. The resulting voltage feedback resolution of the present setup is readily in the \textmu V range \cite{Luepke2015,Luepke2017}.\\

\section{Noise analysis} \label{noise}
\begin{figure}
\includegraphics{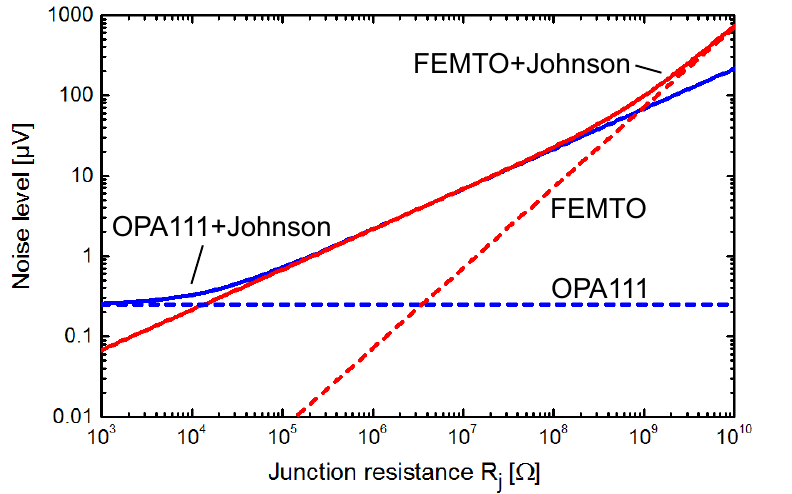}
\caption{Detector noise $V_{\rm detector}$ of voltage follower (dashed blue line) and voltage feedback (dashed red line) circuitry and resulting respective noise limit $V_{\rm limit}$ (solid lines) in combination with Johnson noise, as a function of the junction resistance $R_{\rm j}$ for detector bandwidths of $\Delta f=275\rm\,Hz$. For large junction resistances $R_{\rm j}\gtrsim1\rm\,G\Omega$ the noise level of both, the voltage follower and voltage feedback method, is too large for typical four-probe measurements $V\gtrsim100\rm\,$\textmu V. At intermediate junction resistances $100\rm\,k\Omega\lesssim R_{\rm j}\lesssim1\rm\,G\Omega$ the noise performance of both implementations is dominated by the Johnson noise resulting in approximately the same noise level. For small junction resistances $R_{\rm j}\lesssim100\rm\,k\Omega$ the noise level of the voltage follower circuit converges to the fundamental input noise level of the used opamp (dashed blue line). For the voltage feedback, the noise level continuously decreases as the junction resistance is further lowered resulting in a lower noise level compared to the voltage follower.}
\label{3}
\end{figure}
To compare the individual voltage measurement methods, we use in the following a measurement bandwidth of $\Delta f=275\rm\,Hz$, which is smaller than the bandwidth of the used current amplifier and voltage follower circuits, respectively. A general limitation of the voltage measurement accuracy is given by the junction resistance $R_{\rm j}$ between the probes and the sample. On the one hand, the thermal (Johnson) noise of the tip-sample junction is given by \cite{Rozler2008}
\begin{equation}
V_{\rm Johnson}=\sqrt{4k_{\rm B} T \Delta f R_{\rm j}}, \nonumber
\end{equation}
where $k_{\rm B}$ is the Boltzmann constant and $T=300\rm\,K$ is the junction temperature. Additionally, large junction resistances result in a smaller lateral current injected into the sample which decreases the signal-to-noise ratio with respect to the detector noise floor. In the case of the voltage follower circuit (A), the detector noise is given by \cite{Rozler2008}
\begin{equation}
V_{\rm detector}=\widetilde{V}_{\rm detector}\sqrt{\Delta f}, \nonumber
\end{equation}
where $\widetilde{V}_{\rm detector}=15\rm\,nV/\sqrt{Hz}$ is the input noise density of the OPA111. The resulting total voltage noise at the tip-sample junction is given by
\begin{equation}
V_{\rm limit}=\sqrt{V_{\rm Johnson}^2+V_{\rm detector}^2}.  \nonumber
\end{equation}

On the other hand, the detector noise limit of the voltage feedback techniques (B) and (C) is limited by the input current noise of the current amplifier $\widetilde{I}_{\rm detector}(\rm gain=10^9\rm\,V/A)=4.3\rm\,fA/\sqrt{Hz}$. Resulting from the current noise in combination with the junction resistance follows a corresponding voltage noise of \cite{Rozler2008}
\begin{equation}
V_{\rm detector}=\underbrace{\widetilde{I}_{\rm detector}R_{\rm j}}_{\widetilde{V}_{\rm detector}}\sqrt{\Delta f}. \nonumber
\end{equation}
In contrast to the voltage follower circuit, this noise limit depends also on the junction resistance $R_{\rm j}$, which means that for vanishing junction resistance, the voltage noise $V_{\rm limit}$ goes to zero.

Figure \ref{3} shows a plot of the theoretical noise limit of the present voltage follower and voltage feedback circuitry as a function of the junction resistance $R_{\rm j}$ and at $\Delta f=275\rm\,Hz$. As evident in the graph, for large junction resistances $R_{\rm j}\gtrsim1\rm\,G\Omega$ the noise level of both, the voltage follower and feedback method, is too large for typical four-probe measurements $V_{\rm limit}\gtrsim100\rm\,$\textmu V. At intermediate junction resistances $100{\rm\,k\Omega}\lesssim R_{\rm j}\lesssim1\rm\,G\Omega$ the noise level of both implementations is almost identical as it is dominated by the Johnson noise and is in an acceptable range for most applications. For small junction resistances $R_{\rm j}\lesssim100\rm\,k\Omega$, the voltage follower approaches the constant noise limit of the OPA111, while the noise level of the voltage feedback technique further decreases. We conclude that for high accuracy measurements, the voltage feedback technique is superior to the voltage follower implementation.\\

A way to further increase the measurement accuracy is an averaging of the measurement signals over a time interval larger than the measurement bandwidth. Hereby, the statistical noise decreases as $\sim1/\sqrt{t}$ with the measurement duration $t$. However, this is only true if the measurement setup itself is stable. In real measurements, thermal drift and fluctuations in the contacts, e.g. due to electromigration, can have significant influence on $R_{\rm j}$ and thus on the noise in the four-probe measurement. As a result, for $t>1/(2\pi\Delta f)$ it can be expected that the noise will first go down with $1/\sqrt{t}$ due to the increased averaging of the statistical noise before it starts to deviate from this behavior as the measurement time becomes longer than the time constant on which the measurement setup is stable. The latter can largely depend on the sample under investigation, the exact measurement setup, the tip material and the voltage measurement method used. In typical four-tip setups, the tips are mounted under $45^{\circ}$ angle with respect to the sample surface, such that contacting the tip to the sample surface results in a flexible junction between the sample and the elastic tip. As a result, in direct contact, drift does not have such a significant effect on the four-probe measurement as the junction resistance does not change significantly during typical measurement durations. In the alternating feedback technique, on the other hand, the separation between voltage sensing tip and sample is continuously controlled by the alternating feedback, such that the drift for all of the present techniques should be comparable.\\

To determine the noise level of the full four-probe $I/V$ measurement, in principle both the noise of the measured injected current $I$ and the voltage drop $V$ have to be considered. However, the Johnson noise occurring at the current injecting contacts correspond to fluctuations in the contact resistances which do not play a role in four-probe measurements. As a result, the error in the measurement of the injected current is only determined by the detector noise which in the present setup is below 
\begin{align}
I_{\rm detector}&=\widetilde{I}_{\rm detector}(\rm gain=10^6\rm\,V/A)\cdot\sqrt{\Delta f}\approx2.2\rm\,pA.\nonumber
\end{align}
As typical injected currents are at least $\sim\rm nA$, the voltage measurement noise is the dominant source of noise in the four-probe measurements.

\section{Four-probe measurements} \label{results}
In the four-probe measurements, we record a fixed number of data points (e.g. 1000) for each $I/V$ curve and record the voltage drop for each increment of injected current. The resulting four-probe resistance $R_{\rm 4P}$ is determined by a linear fit to the data around $0\rm\,A$ and for a two-dimensional conductor is independent of the tip spacing when measured with an equidistant tip-tip separation \cite{Schroder2006}. The corresponding sample sheet conductivity is
\begin{equation}
\sigma_{\rm 2D}=\frac{\ln(2)}{R_{\rm 4P}\pi}. \nonumber
\end{equation}
As errors in the positioning of the tips can have a large influence on the measured conductivity \cite{Perkins2013}, for the comparison of the different techniques we have performed each measurement without repositioning the tips between the individual measurements on each sample. As a result, positioning errors can be excluded from the comparison of the different measurements as they only influence the absolute resistance measured but not the relative errors which we will compare in the following. 

\subsection{BiSbTe$_3$}

\begin{figure}
\includegraphics{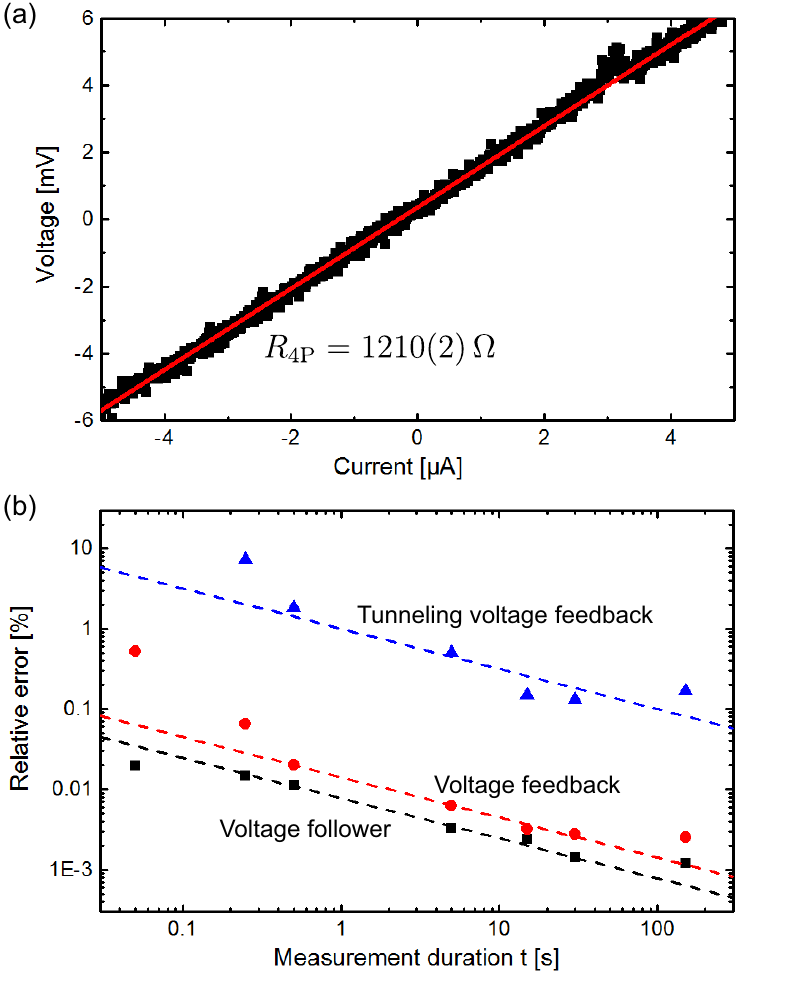}
\caption{(a) Equidistant four-probe $I/V$ of \BiTe measured by tunneling voltage feedback (measurement duration $t=15\rm\,s$) with a linear fit. (b) Error of the resistance measurement relative to the measured resistance for the three different measurement techniques and for different measuring durations on BiSbTe$_3$. The dashed lines correspond to the expected behavior of statistical noise $\sim1/\sqrt{t}$. For shorter measurement durations, the voltage feedback does not fully converge, resulting in additional measurement errors while for longer measuring durations ($>30\rm\,s$) variations in the tip contacts increase the error in the data.}
\label{4}
\end{figure}

We first demonstrate four-probe measurements using the different voltage measurement techniques on a $10\rm\,nm$ thin film of BiSbTe$_3$. The sample was grown on Si(111) by molecular-beam epitaxy and was exposed to air prior to the analysis in the four-tip STM chamber \cite{Luepke2017}. The result of an equidistant four-probe measurement on the BiSbTe$_3$ film using the tunneling voltage feedback technique and a measurement duration $t=15\rm\,s$ is shown in Fig. \ref{4} (a). The linear fit results in $R_{\rm 4P}=1210(2)\rm\,\Omega$ and therefore $\sigma_{\rm 2D}=0.1823(3)\rm\,mS/\ensuremath{\Box}$, corresponding to a relative error of 0.17\%. The comparison of the relative errors for the different measurement durations and techniques is shown in Fig. \ref{4} (b). We find that the data of the tunneling voltage feedback measurements shows generally a higher noise level compared to the other two methods, which have comparable noise levels. We attribute this finding to the larger junction resistance of the tunneling contact in comparison to the other measurement methods in combination with smaller averaging duration during the measurement as result of the alternating feedback. In the same graph, we have also plotted dashed lines corresponding to $a\sim1/\sqrt{t}$ behavior for each of the three methods which represents the theoretical dependence of statistical noise on the tunneling resistance and which describe the data quite well. We attribute this observation to the fact that the sample is quite soft and therefore the contacts of the tips with the sample are rather stable such that the combined Johnson and amplifier noise is the dominant source of noise for $0.5\rm\,s\leq t\leq30\rm\,s$.  
For $t<0.5\rm\,s$, the data points of the voltage feedback measurement deviate from the graph of the statistical noise. In this range, due to the short measurement duration the voltage feedback loop does not fully converge to each new value of $V_{\rm local}$ as the lateral current is swept, resulting in additional errors in the measurements. For $t>30\rm\,s$, we find that measurement instabilities begin to increase the measurement errors above the expected statistical error behavior. We estimate the junction resistances of the current injecting tips from the difference of the measured 2D sheet conductivity in comparison to the current/voltage characteristics at the current injecting tips to be $R_{\rm j}\approx9\rm\,k\Omega$. The junction resistance for the tunneling voltage feedback is $R_{\rm j}\approx40\rm\,M\Omega$

\subsection{\Si} 

\begin{figure}
\includegraphics{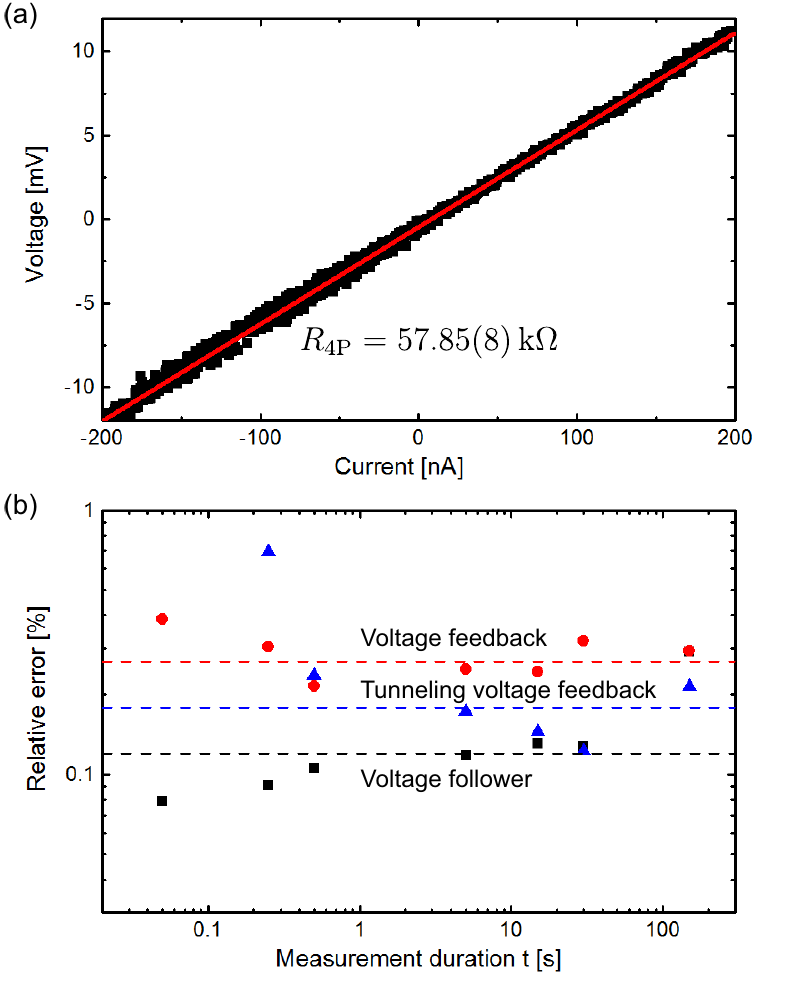}
\caption{(a) Equidistant four-probe $I/V$ of \Si measured by tunneling voltage feedback (measurement duration $t=15\rm\,s$) with a linear fit. (b) Relative noise of the different measurement techniques for different measuring durations $t$ on Si$(111)-(7\times7)$. The dashed lines correspond to the average constant noise levels of the three measurement techniques. For shorter measurement durations the limited bandwidth of the voltage feedback based techniques increases the measurement error.}
\label{5}
\end{figure}

We further applied the different four-probe measurement techniques to the \Si surface. The transport properties of this surface have been under recent discussion because earlier results on its surface conductivity varied largely as a result of numerous different measurement and sample preparation procedures \cite{Hofmann2009}. One way to measure the surface conductivity is to determine the two-dimensional contribution in distance-dependent four-probe measurements \cite{Dangelo2009,Martins2014,Luepke2015,Just2017}. Fig. \ref{5} (a) shows a typical four-probe measurement on this surface in tunneling voltage feedback mode. The linear fit results in $R_{\rm 4P}=57.85(8)\rm\,k\Omega$ corresponding to $\sigma_{\rm 2D}=3.814(5)\cdot10^{-6}\rm\,S/\ensuremath{\Box}$. This result is in general agreement with previous measurements on the same samples which used the voltage follower method \cite{Just2015}. For a more detailed comparison, Fig. \ref{5} (b) shows the noise levels for the three different techniques and at varying measurement duration on the same sample without repositioning the tips. While there are deviations for short measurement durations $t<0.5\rm\,s$ which are again the result of the voltage feedback loop not fully converging, especially in tunneling contact, we find that noise level of the three different techniques approximately coincide for measurement durations $t\geq0.5\rm\,s$. In this range, the data shows rather constant noise levels for all three techniques, which means that the larger amount of noise resulting from the tunneling contact of the voltage probe in contrast to the voltage measurement in direct contact is insignificant under the present measurement conditions. We explain this finding to be due to large fluctuations in the tip contacts during the measurements resulting from the hard sample surface and non-linearities in the four-probe $I/V$ curves which dominate the noise in the measurements. We estimate the junction resistance of the current injecting tips in contact to be $R_{\rm j}\approx200\rm\,k\Omega$. In comparison, the junction resistance of the tunneling tip in the present case is $R_{\rm j}\approx20\rm\,M\Omega$.

\section{Discussion} \label{appl}
\begin{figure}
\includegraphics{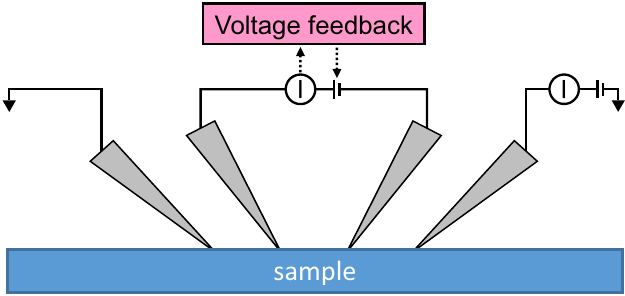}
\caption{Schematic of the most simple four-point probe measurement setup using the voltage feedback technique. The current through the sample is injected between a tip connected to a biased current sensor and a tip on ground potential. The voltage drop across the sample is determined by the two inner tips which are connected to a biased current sensor which nulls the current between the voltage sensing tips. The voltage required to compensate the current through the tips corresponds to the voltage drop across the sample surface between the two contact points of the voltage sensing tips.}
\label{6}
\end{figure}
The most simple measurement setup of a four-point measurement resulting from the voltage feedback method is shown in Fig. \ref{6}. Here, only two current measurement units are required, one to inject the lateral current to another tip on ground potential and the other to null the current between the two voltage sensing tips. The resulting voltage at which no current flows through the voltage sensing tips corresponds to the voltage drop across the sample surface between the respective contact points of the voltage sensing tips.

The present approach further allows to have all four tips required for the four-probe measurement in tunneling contact to the sample, which enables truly non-invasive four-probe measurements. For this application the only additional requirement is that the sample needs to be on floating potential during the potential measurement and on fixed potential for the topography feedbacks of the tips. While we have not yet implemented this approach into the present setup, yet, it can be realized e.g. by switching of a relay connecting the sample to ground synchronous with the topography feedback. Note that this relay would not result in much additional noise in the actual four-probe measurement because it is not located in any of the current paths of the actual measurement. An alternative would again be the use of an AC component for the topography feedback and a DC component for the actual four-probe measurement. In this realization, to decouple the DC conductivity measurement, the sample could be connected via a capacitor to a sine generator, which provides the AC signal. We further propose that it is also possible to use a different mechanism for the topography feedback, e.g. atomic force microscopy (AFM) based feedback, to maintain a constant tip-sample distance while the potential measurement is continuously performed via the tunneling current. The advantage of an AFM feedback to control the tip heights is that the sample can be on floating potential throughout the entire measurement\footnote{A corresponding patent has been filed}.

A general advantage of non-invasive four-probe measurements is that the damaging of the samples and tips is greatly reduced compared to conventional four-probe measurements. This is important because a change in the sample surface structure can significantly alter its transport properties. While for the measurements shown in the present manuscript the influence of tip contacting on the measured conductivity is negligible because of the spatial extent of the samples, this aspect becomes more important for smaller and more fragile nanostructures, e.g. monoatomic wires. Furthermore, for samples in the ballistic transport regime especially the invasiveness of the voltage probes can have a significant influence on the measured transport properties \cite{Baringhaus2014}. Using the tunneling voltage feedback technique one can precisely control the invasiveness of these probes. Concerning the tip wear, conventional four-probe measurements lead to a blunting and bending of the tips such that they have to be exchanged regularly. In contrast, four-probe measurements in tunneling contact prevent much of this wear and increase the tip lifetime significantly.

\section{Conclusion} \label{conc}
We have demonstrated a multi-point probe implementation which uses equivalent minimalistic current sensors at each probe and where voltage measurements are performed by a software voltage feedback loop. The resulting setup performance, under typical experimental conditions, is comparable to previous voltage follower setups, but with less complicated electronics. As we have used here the same electronics for the comparison of the different measurement methods, in the final application the noise level of the voltage feedback techniques will be even lower than demonstrated here, when the electronics are reduced to only the required biased current amplifiers. Nevertheless, we find that all of the above measurement errors are small ($\leq1\%$) compared to other typical error sources in four-probe measurements, especially the positioning errors of the tips which result in relative errors in the measured conductivity of $\sim10\%$ \cite{Just2015,Perkins2013}. The implemented tunneling voltage feedback technique further allows the non-invasive characterization of sample transport properties, compatible with vacuum conditions, low temperatures, gating and magnetic fields which makes it a powerful tool for future studies.\\

\section*{ACKNOLEDGEMENTS}
We would like to thank the group of G. Mussler from Peter Gr\"unberg Institut (PGI-9), Forschungszentrum J\"ulich, for the preparation of the BiSbTe$_3$ sample.


%

\end{document}